\documentclass[twocolumn,english,aps,floatfix,superscriptaddress,showpacs,amsfonts,amssymb,superscriptaddress,nofootinbib]{revtex4}
\usepackage{color} 
\usepackage{slashed}
   
\usepackage{graphicx}
\usepackage{amsmath}
\usepackage{latexsym} 
\usepackage{epstopdf}
\usepackage{amsmath}
\usepackage{amssymb,amsmath}
\usepackage{multirow}
\usepackage{mathtools}
\newcommand{\beq}{\begin{equation}}
\newcommand{\eeq}{\end{equation}}

\newcommand{\be}{\begin{equation}}
\newcommand{\ee}{\end{equation}}

\newcommand{\spinor}[2]{
\left(\begin{matrix} #1 \\  #2 \end{matrix}\right)}

\begin{document}

\title{Probing  Klein tunneling through quantum quenches}

\author{Leda Bucciantini}
\affiliation{Dipartimento di Fisica dell'Universit\`a di Pisa and INFN - 56127 Pisa, Italy}
\affiliation{SISSA, via Bonomea 265, 34136 Trieste, Italy}
\author{Spyros Sotiriadis}
\affiliation{ SISSA and INFN, Sezione di Trieste, Trieste, Italy}
\author{Tommaso Macr\`i}
\affiliation{Departamento de F\'isica Te\'orica e Experimental, Universidade Federal do Rio Grande do Norte, 59072-970 Natal-RN,Brazil}
\affiliation{International Institute of Physics, 59078-400 Natal-RN, Brazil}

\begin{abstract}
We study the interplay between an inhomogeneous quantum quench of the external potential in a system of relativistic fermions in one dimension and the well-known Klein tunneling. We find that the large time evolution is characterized by particle production at a constant rate which we derive analytically. The produced particles can be physically interpreted according to a semiclassical picture and the state reached in the long time limit can be classified  as a non-equilibrium-steady-state. Such a quantum quench can be used in order to observe macroscopic effects of  Klein tunneling in transport, with potential implementations with current experimental setups.
\end{abstract}
\maketitle

\section{Introduction}

Klein tunneling  \cite{klein} refers to the fact that an incoming relativistic electron can penetrate a potential barrier of height $V$ greater than twice the electron's rest mass $m$ if the electron's energy lies between $m$ and $V-m$\footnote{Throughout the paper we will use natural units $\hbar=c=1$.}.
In particular, keeping the energy of the incident particle fixed and increasing the height of the barrier, the transmission coefficient does not decay, as one would probably expect and as it actually happens in the bosonic analogue \cite{Hol}. 
The key feature of this phenomenon is the fact that the electrons are relativistic and in fact the necessary tool to correctly explain it is quantum field theory \cite{saut, hund, Feynman}. After its explanation (see \cite{Hol, Dombey, Su} for more recent reviews), the Klein tunneling phenomenon gained much relevance along with the discovery of particle-antiparticle production from a strong enough potential \cite{Schwinger}, vacuum polarization effects \cite{Rafelski} and black hole evaporation due to the creation of particle-antiparticle pairs near the event horizon (Hawking radiation) \cite{Hawking}.

This and other peculiar effects of the Dirac equation, like the \emph{Zitterbewegung} \cite{Shr}, although representing key phenomena to understand relativistic quantum effects, have proven difficult to observe experimentally. For instance, the observation of Klein tunneling requires a potential drop of the order of the fermion mass $m$  over the Compton length $1/m$ which yields an enormous electric field \cite{Greiner,Grib}, thus making the effect relevant only for very exotic situations \cite{Greiner, Grib, Page}. These difficulties stimulated a great interest for the simulations of relativistic quantum systems with condensed matter setups in the laboratory \cite{Blatt}.
The pioneering study of graphene \cite{Novoselov, Pereira} is certainly the primary example. More in general, using recently developed techniques, like ultracold atoms in optical lattices \cite{review_cold_atoms}, ions \cite{ions} or photonic systems \cite{photonic_systems}, it is possible to device highly tunable systems which allow for the preparation and detection of a great variety of many-body phenomena. Examples include the simulation of black holes in Bose-Einstein condensates \cite{Garay}, Dirac equation in various dimensions \cite{Lamata}. Recently Klein tunneling was also simulated in single ion traps \cite{gerritsma_2011}.

In this paper we study another physical situation where the features of Klein tunneling emerge via 
a sudden quench of the external potential for a system of relativistic one-dimensional Dirac 
fermions from a constant zero value to a step-like profile.

The sudden quench protocol \cite{CC} consists of the following three steps: (1) prepare a system in a pure state, usually the ground state of a pre-quench Hamiltonian; (2) at a certain time, namely $t=0$, suddenly vary one of the parameters of the Hamiltonian; (3) from that time on, let the system evolve unitarily, i.e. without connection or dissipation to the external environment, according to the post-quench Hamiltonian which does not commute with the pre-quench one. Being the pre-quench and post-quench Hamiltonian not mutually commuting guarantees that the time evolution starts from an initial out-of-equilibrium configuration. 

This is the simplest way of driving a many body quantum system out-of-equilibrium and has the advantage that analytical calculations of the long time limit of observables can be carried out in many cases \cite{QQ_analytics,review_quench}. When the pre-quench Hamiltonian is inhomogeneous, it is also possible to study transport properties of the system \cite{transport}. It is important to highlight that many of these recent theoretical investigations are triggered by several experimental realizations of quantum quenches \cite{exp} in engineered low dimensional quantum systems in ultracold atoms \cite{review_cold_atoms}, which allow to have coherent dynamics for much longer times than with usual solid state systems.

The reason why  Klein tunneling can be linked to the quantum quench scenario relies on the characterization of the initial state in which the system is prepared before the quench in terms of the post-quench Hamiltonian. It is a well-established semiclassical picture \cite{CC,semicl_ex} that by performing a global quench on the system (i.e. changing some parameter in the whole system) we inject an extensive amount of energy and the initial state can be thought of as containing an extensive number of quasi-particle excitations which, after the quench, start spreading ballistically. In the present case we show that, as a consequence of quenching the potential in the one dimensional Dirac equation from a constant zero value to a step-like profile with height $V>2m$, at large times there is particle production at a constant rate, which we derive analytically. Our rigorous field theoretical derivation also provides a solid theoretical framework for previous numerical investigations of particle production \cite{Krekora, Cheng} and single particle quantum mechanical calculations \cite{Dombey} that perfectly agree with our findings.

The manuscript is organized as follows: in section (\ref{two}) and (\ref{three}) we carefully re-derive the energy eigenstates that are  solutions of the Dirac equation in a homogeneous potential and in an inhomogeneous one of a step-like form, pointing out the differences depending on the height of the potential. In section (\ref{four}) we introduce the quench  of the potential from   $V=0$  to $V(x)=V \Theta(x)$, in particular we derive the overlaps between the pre-quench and post-quench eigenfunctions; then we analytically compute the number of pre-quench particles produced through the scattering process at large times and give a physical interpretation to it.

\section{One-dimensional Dirac fermions in presence of potential step}\label{two}

\subsection{Homogeneous case}

We consider a one-dimensional system of relativistic fermions of mass $m$ in an external potential $V(x)$. This is described by the Dirac equation 
$(i\slashed\partial - m - \gamma_0 V(x) ) \psi(x,t) = 0$, 
which, using the representation $\gamma_0=\sigma_3, \gamma_1=i\sigma_1$, reads
\be
((i\partial_t-V(x))\sigma_3 - \sigma_1 \partial_x - m) \psi(x,t) = 0, \label{Dirac_hom_0}
\ee
where $\sigma_{1,2,3}$ are the Pauli matrices. 
The energy eigenstates satisfy the time-independent form of the Dirac equation 
\be
[(E-V(x)) \sigma_3 - \sigma_1 \partial_x - m] \psi(E;x) = 0, \label{Dirac_hom}
\ee
where $\psi(E;x)$ is an eigenstate of energy $E$.

\begin{figure}[h!]
\begin{center}
\includegraphics[clip,width=0.9\linewidth]{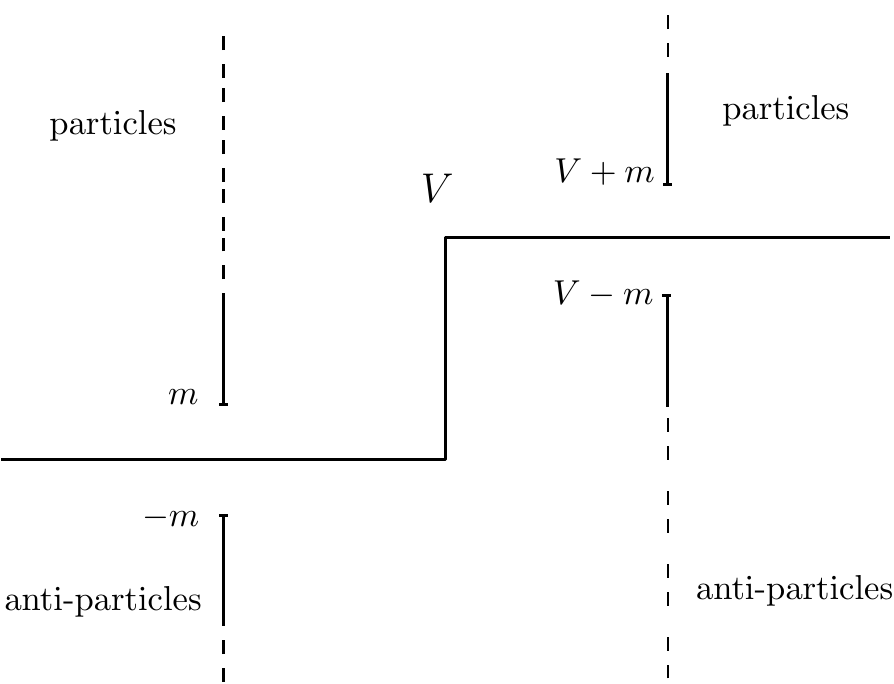}
\caption{\label{fig:fig2} Energy windows for the eigenfunctions in the homogeneous case, with distinction between the particle and antiparticle solutions. On the left, the ones for the $V=0$ case, on the right the ones for the $V$ case.}
\end{center}
\end{figure}

In the absence of a potential, i.e. when $V(x)=0$, the solutions to the above equation are 
\begin{align}
& \psi(E;x)=u^\pm(E;x) \equiv \sqrt{\frac{E+m}{4\pi k_{E}}} \spinor{i}{\frac{\pm{k_{E}}}{ E+m}} 
e^{\pm i{k_{E}} x}, 
\end{align}
where ${k_{E}}=\sqrt{E^2-m^2}$ with $|E|\geq m$. Positive energy solutions $E\geq+m$ describe particles while negative energy solutions $E\leq-m$ describe antiparticles. For each energy $E\geq+m$ ($E\leq-m$) there are two independent solutions corresponding to left or right moving particles (right or left moving antiparticles, respectively). 
Note that the above solutions are orthonormalized in infinite volume i.e. 
\be
\int\limits_{-\infty}^{+\infty} dx\; \left(u^s(E;x)\right)^\dagger u^{s'}(E';x) = \delta_{ss'}\delta(E-{E'})
\ee
If the potential $V(x)$ is homogeneous i.e. $V(x)=V_0$, then the solutions are $\psi(E;x)=u^\pm(E-V_0;x)$; consequently particles are described by solutions with $E>V+m$ while antiparticles by solutions with $E<V-m$ (see Fig.~\ref{fig:fig2}).

\subsection{Inhomogeneous case}

Assuming now a potential step $V(x)=V\Theta(x)$ (with $V>0$) the new solutions can be found by matching together at the origin the homogeneous solutions in the positive and the negative semi-axes with the corresponding values of the potential, keeping in mind that solutions with imaginary wavenumbers are also acceptable in the semi-axes as long as they decay exponentially at large distances. The matching is prescribed by the continuity condition $\psi(E;0^-)=\psi(E;0^+)$ at the origin. Depending on the value of $E$, the wavenumbers $q\equiv k_{E-V}$ on the left and $p\equiv k_{E}$ on the right may be both real, one real and the other imaginary or both imaginary. We therefore distinguish the following cases, which are depicted for clarity in Fig.~\ref{fig:fig4}:

\begin{figure}[h!]
\begin{center}
\includegraphics[clip,width=1.0\linewidth]{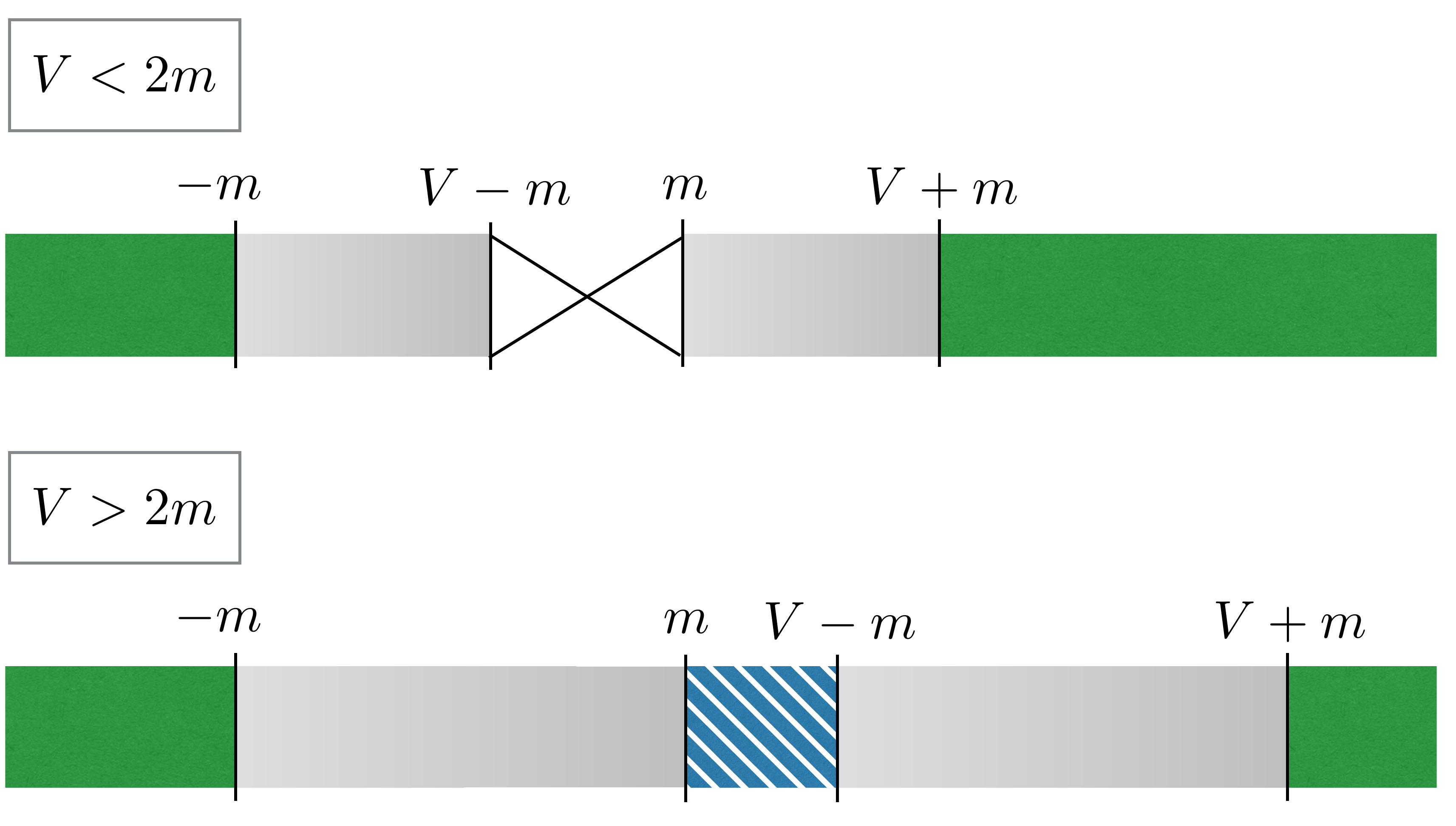}
\caption{\label{fig:fig4} Energy windows for the different kinds of eigenfunctions in the inhomogeneous potential case. Grey zones denote totally reflecting solutions, green ones plane wave solutions. The crossed one, only present in the $V<2m$ case, represents the not allowed region, while the dashed one, present only in the $V>2m$ case, represents the Klein Zone.}
\end{center}
\end{figure}

\begin{figure}[h!]
\begin{center}
\includegraphics[clip,width=0.9\linewidth]{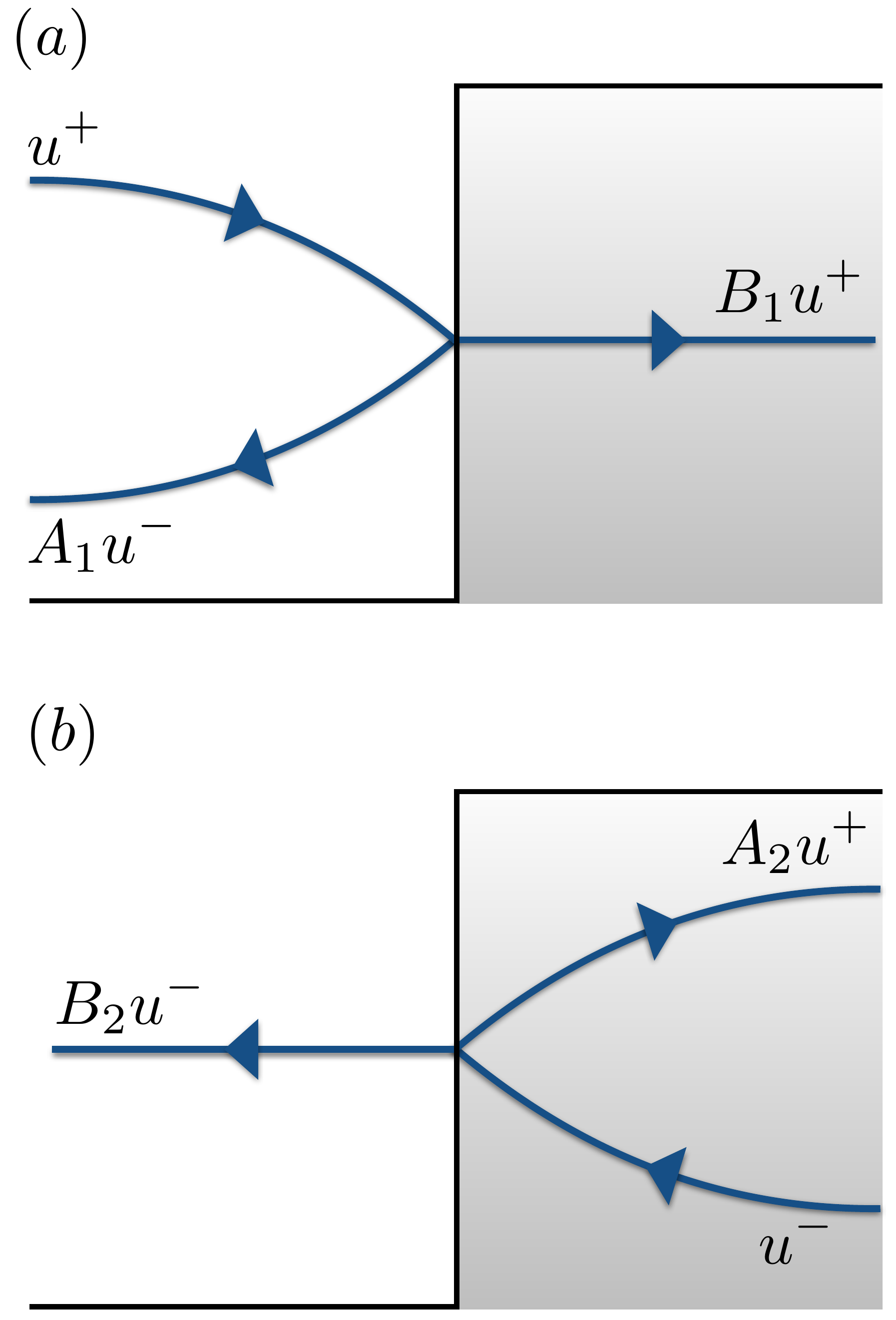}
\caption{\label{fig:fig5} Pictorial representation of eigenfunctions (\ref{v1}) (see (a)) and (\ref{v2}) (see (b)) for $E>V+m$.}
\end{center}
\end{figure}

1) $V<2m$: 
In this case there exist three different energy windows. For energies $E\in(-\infty,-m]\cup[V+m,+\infty)$
both $q$ and $p$ are real and the solutions correspond to plane waves on both sides of the step. They represent particles or antiparticles that cross the potential step and are partially reflected and partially transmitted to the opposite side. Since the incident particle or antiparticle may come from either side of the origin, the solutions are doubly degenerate. For energies $E\in(-m,V-m)\cup(m,V+m)$ instead, exactly one of $q$ and $p$ is real while the other is imaginary, therefore the solutions correspond to plane waves on one side of the step and exponentially damped waves on the other. They represent particles or antiparticles that are totally reflected by the potential step and are non-degenerate since, in order to have exponential damping on one side, the incident plane wave can come only from the other side. Lastly, for energies $E\in[V-m,m]$ both $q$ and $p$ are imaginary and so the associated solutions, if there existed, would be bound states. However the continuity condition is not satisfied for any value of the energy in this window, so there do not exist any solutions of this type.

2) $V> 2m$: 
In this case the energy windows are as follows: solutions are plane waves on both sides for energies $E\in(-\infty,-m]\cup[m,V-m]\cup[V+m,+\infty)$,
while for all other energies $E\in(-m,m)\cup(V-m,V+m)$ the solutions are totally reflecting. 

Note that, as already known from standard quantum field theory, in the ground state the infinite set of negative energy eigenstates should be considered as occupied up to some energy level, the Fermi sea level $E_F$, in order for the energy spectrum to be bounded from below, so that the theory makes physical sense. Excitations above the ground state are either occupied eigenstates with energy $E>E_F$ or unoccupied eigenstates with energy $E<E_F$ (hole or antiparticle excitations). $E_F$ is typically chosen to be zero, however in the present problem we would rather let ourselves free for the moment.

Let us first focus on the case $V<2m$. The (doubly degenerate) reflecting-transmitting solutions are given by
\begin{align}
& v_1(E;x) =
\begin{cases}
u^+(E;x) + A_1 u^-(E;x) , & \quad x<0\\
 B_1 u^+(E-V;x) , & \quad x>0\\
\end{cases} \label{v1} \\
&  v_2(E;x) = 
\begin{cases}
B_2 u^-(E;x) , &  x<0\\
u^-(E-V;x) + A_2 u^+(E-V;x) , & x>0\\
\end{cases} \label{v2}
\end{align}
where, using the matching condition, the coefficients $A_{1,2}$ and $B_{1,2}$ are found to be 
\begin{align}
& A_1 = \frac{1-\kappa}{1+\kappa} , && B_1 = \frac{2\sqrt{\kappa}}{1+\kappa} , \\
& A_2 = -A_1 , && B_2 = B_1 , \label{AB}
\end{align}
with 
\be
\kappa \equiv  \frac q p \frac{(E+m)}{(E-V+m)},  \label{kappa} 
\ee
and 
\begin{align}
q&=k_{E-V}=\sqrt{(E-V)^2-m^2}, \\
p&=k_{E}=\sqrt{E^2-m^2}. 
\end{align}

The reflection $R$ and transmission $T$ probability coefficients for both $v_{1,2}$ are given by 
\begin{align}
R&=|A|^2 = \left( \frac{1-\kappa}{1+\kappa}\right)^2, \label{R} \\
T&=|B|^2 = \frac{4|\kappa|}{(1+\kappa)^2} . \label{T}
\end{align}
For energies $E\in(-\infty,-m]\cup[V+m,+\infty)$, the parameter $\kappa$ is positive,  hence $A_i$ and $B_i$ are real and therefore $R+T=1$. This relation expresses the conservation of probability currents.\\
 The physical meaning of (\ref{v1}) and (\ref{v2}) for  $E> V+m$  is explained respectively in Fig.~\ref{fig:fig5}(a) and (b). For  $E> V+m$  the eigenfunctions represent particles both in (\ref{v1}) and (\ref{v2}), so the group velocity $v_g \equiv \frac{\partial E(k)}{\partial k}$, indicated by the arrow, is in the same direction as the momentum (indicated by the index $\pm$ of the $u$ function). In Fig.~\ref{fig:fig5}(a), for $x<0$ $u^+$ and $u^-$ are respectively the incoming (from the left) and reflected (to the left) particles, while for $x>0$ $u^+$ is the transmitted (to the right) particle. In  Fig.~\ref{fig:fig5}(b),  for $x<0$, $u^-$ is the transmitted (to  the left) particle, while for $x>0$ $u^-$ and $u^+$ are respectively the incoming (from the right) and reflected (to the right)   particle.

The totally reflecting solutions are given by the same expressions, more precisely by (\ref{v1}) if $E\in(m,V+m)$ and by (\ref{v2}) if $E\in(-m,V-m)$ with the coefficients $A_i$ and $B_i$ still given by (\ref{AB}) but now being complex. $A_i$ in particular is unitary (i.e. $|A_i|^2=1$) since $\kappa$ is imaginary. The reflection and transmission coefficients are in this case $R=1$ and $T=0$. 
Note that these are non-degenerate solutions: the first solution corresponds to a particle incident from the left and exists for $E\in(m,V+m)$, while the second corresponds to an antiparticle incident from the right (with coefficient $A_2$) and exists for $E\in(-m,V-m)$. The non-degeneracy is because it is only one of the two complex-conjugate imaginary wavenumbers $\pm k_{E-V}=\pm i|k_{E-V}|$ (or $\pm k_E = \pm i|k_E|$) that results in a wavefunction that is exponentially decaying at large distances $x\to+\infty$ (or $x\to-\infty$ respectively). 

It can be verified that the above eigenstates are by construction orthonormalized, since $u^\pm$ are orthonormalized too. In particular the pair of degenerate states (\ref{v1}) and (\ref{v2}) is chosen in such a way that they are orthogonal to each other
\begin{align}
& \int\limits_{-\infty}^{+\infty} dx\; \left(v_1(E;x)\right)^\dagger v_2(E';x) = \nonumber \\
& \qquad \tfrac12 (A_1^* B_2 + B_1^* A_2) \delta(E-{E'}) = 0, \label{orthg}
\end{align}
since $A_1$ and $B_1$ are real. 
Normalization is ensured by the probability conservation relation $R+T=1$,
\begin{align}
& \int\limits_{-\infty}^{+\infty} dx\; \left(v_s(E;x)\right)^\dagger v_s(E';x) =  \nonumber \\
& \tfrac12(1+R+T)\delta(E-{E'})=\delta(E-{E'}), \quad s=1,2. \label{othnorm}
\end{align}

\section{Klein tunneling}\label{three}

We now turn our attention to the case $V>2m$. While it is still true that for energies $E\in(-\infty,-m]\cup[V+m,+\infty)$, the parameter $\kappa$ is positive and $R+T=1$, this is no longer true within the energy window $E\in(m,V-m)$, which we will call from now on \emph{`Klein zone'}. This is because the parameter $\kappa$ becomes negative and so $R$ and $T$ as defined in (\ref{R}), (\ref{T}), now satisfy $R-T=1$. The probability conservation seems then to be violated since $R+T\neq 1$. Moreover increasing the potential step $V$ the parameter $\kappa$, for energies in the middle of the zone $(m,V-m)$, tends to -1 so that $T$ tends to unity.

\begin{figure*}
\begin{center}
\includegraphics[width=0.75\textwidth]{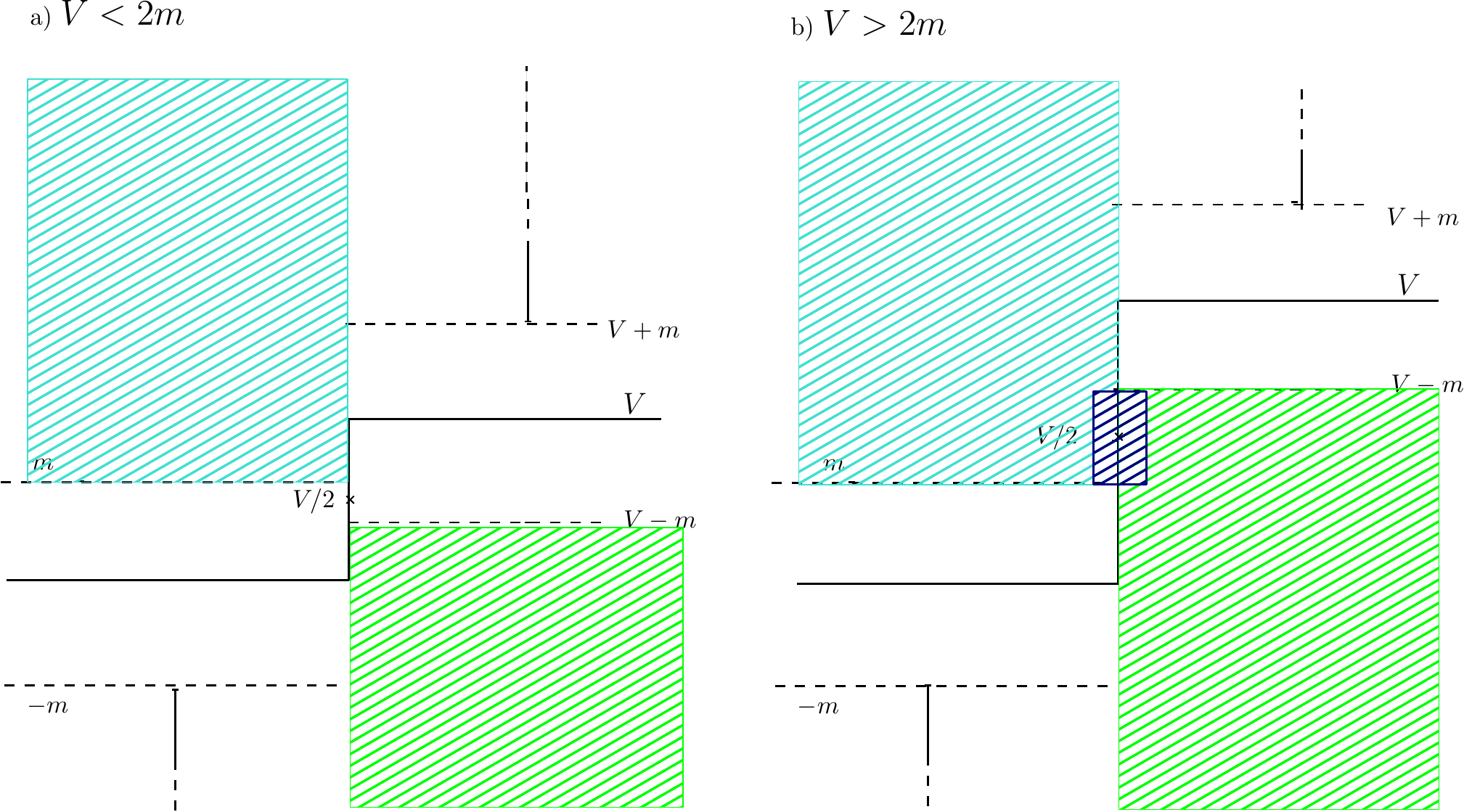}
\caption{\label{fig:fig3b}Absence (a) and presence (b) of Klein tunneling respectively for the cases $V<2m$ and $V>2m$. We highlight the energy window for particles in $x<0$ (light-blue), for antiparticles in $x>0$ (green). Only for $V>2m$ there is a non vanishing overlap between the two (the blue region).}
\end{center}
\end{figure*}

As first clarified by  Sauter \cite{saut} and Hund \cite{hund}, this phenomenon is paradoxical only as long as it is understood in a one-particle quantum mechanical framework but there is no room for any paradox within second quantized field theory, after taking into account the possibility of particle-antiparticle production \cite{Rafelski, Hol}. When an external potential is high enough it allows for the spontaneous production of particles-antiparticles pairs \cite{Greiner, Grib}; the barrier is repulsive for electrons but attractive for positrons. Then  Klein tunneling arises when fermion states outside the barrier have energy that matches the energy of the antiparticles inside the barrier \cite{Su, Dombey}. Since electrons in $x<0$ have energies larger than $m$ while positrons in $x>0$ have energies smaller than $V-m$, a non vanishing overlap between these two regions is present only for $V>2m$ and actually represents the Klein zone (see Fig.~\ref{fig:fig3b}). \\
As a consequence, for all energy zones where $\kappa>0$ the eigenfunctions maintain the same form as in the case $V<2m$. On the other hand, for $V>2m$ and $\kappa<0$, the coefficients $A_i$ are real but $B_i$ are imaginary, so that neither orthogonality (\ref{orthg}) nor (\ref{othnorm}) holds for the  states (\ref{v1}) and (\ref{v2}).

The correct ones can be found noticing that,  for the state $v_1$ to describe a scattering process with a transmitted (i.e. right-moving) antiparticle on the right, the direction of momentum on that side must be flipped. Similarly, for $v_2$ both directions of the plane waves on the right should be flipped (in which way the amplitude of the incoming particle is still equal to 1). We thus define, in the Klein zone, the new pair of degenerate states
\begin{align}
& v'_1(E;x) =
\begin{cases}
u^+(E;x) + A'_1 u^-(E;x) , & \quad x<0\\
 B'_1 u^-(E-V;x) , & \quad x>0\\
\end{cases} \label{v'1} \\
&  v'_2(E;x) = 
\begin{cases}
B'_2 u^-(E;x) , &  x<0\\
u^+(E-V;x) + A'_2 u^-(E-V;x) , & x>0\\
\end{cases} \label{v'2}
\end{align}
and using the matching condition we find that the new coefficients $A'_{1,2}$ and $B'_{1,2}$ are
\begin{align}
& A'_1 = \frac{1+\kappa}{1-\kappa} = \frac{1-\kappa'}{1+\kappa'}, && B'_1 = \frac{2\sqrt{\kappa}}{1-\kappa} =\frac{2i\sqrt{\kappa'}}{1+\kappa'} , \\
& A'_2 = -A'_1 , && B'_2 = - B'_1 , \label{AB'}
\end{align}
where
\be
\kappa' \equiv - \kappa = - \frac q p \frac{(E+m)}{(E-V+m)} > 0 . \label{kappa2} 
\ee
Now the reflection $R$ and transmission $T$ probability coefficients, defined as the reflected-to-incoming and transmitted-to-incoming probability current ratios, are given by
\begin{align}
R&=|A'|^2 = \left( \frac{1-\kappa'}{1+\kappa'}\right)^2, \label{R'} \\
T&=|B'|^2 = \frac{4|\kappa'|}{(1+\kappa')^2} ,  \label{T'} \\
& \qquad \text{ for energies in the Klein zone,} \nonumber
\end{align}
and since $\kappa'>0$, the probability conservation relation $R+T=1$ is recovered.

The new states (\ref{v'1}) and (\ref{v'2}) are orthogonal to each other
\begin{align}
& \int\limits_{-\infty}^{+\infty} dx\; \left(v'_1(E;x)\right)^\dagger v'_2(E';x) = \nonumber \\
& \qquad \tfrac12 ({A'}_1^* B'_2 + {B'}_1^* A'_2) \delta(E-{E'}) = 0, \label{orthg2}
\end{align}
after taking into account  (\ref{AB'}) and  that  $B'_1$ is imaginary. Each of these states is normalized to unit
\begin{align}
& \int\limits_{-\infty}^{+\infty} dx\; \left(v'_s(E;x)\right)^\dagger v'_s(E';x) =  \nonumber \\
& \qquad \tfrac12(1+R+T)\delta(E-{E'})=\delta(E-{E'}) , \quad s=1,2. \label{othnorm2}
\end{align}

It should be stressed that redefining the eigenstates for the Klein zone as above is a necessity rather than an arbitrary choice: indeed, for the subsequent study orthogonality and normalization are necessary. If we chose to keep one of the eigenstates, say $v_2$, in the same form as in the other zones, we would have to multiply it by $1/A_2$ in order to make it normalized in the Klein zone, thus obtaining $v'_2=v_2/A_2$, and next to choose the other eigenstate as the (unique) linear combination of $v_1$ and $v_2$ that is orthogonal to $v'_2$ and normalized too, thus obtaining $v'_1=v_1-(B_1/A_2) v_2$. The new pair of states $v'_1$ and $v'_2$ are then precisely the ones given by (\ref{v'1}) and (\ref{v'2}). 

Notice that since the group velocity of an antiparticle is opposite to its momentum, all eigenstates corresponding to antiparticles should be defined, for any value of the potential, with their momentum signs flipped, if we want them to describe physical scattering processes in which the incident wave has coefficient equal to 1.   When all three plane waves (incoming, reflected and transmitted) correspond to antiparticles (for instance in the case $E<-m$ of (\ref{v1})  and (\ref{v2})) flipping all three momentum signs of the eigenstates does not amend the values and physical significance of their coefficients $A_i,B_i$ or those of $R$ and $T$, neither does it spoil the orthonormalization relation of the two degenerate states. On the contrary, in the Klein zone, since in $x<0$ there are particles while  in $x>0$ antiparticles, only the momenta of the plane waves in $x>0$ must be flipped. For this reason, while it is crucial to redefine the states in the Klein Zone, this is not necessary for all other energy windows and we can keep them in the original form (\ref{v1}) and (\ref{v2}). Hence, as a  pictorial representation for the eigenfunctions (\ref{v'1}) and (\ref{v'2}) in the Klein zone, we can refer to Fig.~\ref{fig:fig5} with the index of $u$ in $x>0$ changed. The arrows represent the correct direction of the group velocity of plane waves.

Except when otherwise stated, in the following we will drop the prime from the notation of expressions (\ref{v'1}) and (\ref{v'2}), that is we redefine $v_{i}$ to be equal to $v'_{i}$ in the Klein zone energy range and similarly for the parameters $A_i$ and $B_i$. 

\section{Quench}\label{four}

We now consider a one-dimensional system of Dirac fermions described by (\ref{Dirac_hom}), initially prepared in the ground state that corresponds to a homogeneous potential $V_0=0$; at $t=0$ the potential is quenched to the inhomogeneous step potential $V(x)=V\Theta(x)$ and the system, from now isolated from the rest, is subject to unitary evolution according to the post-quench Hamiltonian.

The pre-quench Hamiltonian in a second quantized form is
\be
H =\sum_ {\sigma=\pm} \left(\int_{m}^{\infty} dE \, E \,\alpha^{\sigma \dagger}_E \alpha^{\sigma}_E - \int_{-\infty}^{-m} dE \, E \,\beta^{\sigma \dagger}_E \beta^{\sigma}_E \right),
\ee
where $\alpha^\sigma_E$ and $\beta^{\sigma \dagger}_E$ are the electron annihilation operator and positron creation operator respectively, obeying usual anticommutation relation
\be \label{rel_antialpha}
\{ \alpha ^\sigma_E, \alpha^{\dagger\sigma'}_{E'}\}=\delta_{\sigma,\sigma'}\delta(E-E'), \quad \{ \alpha ^\sigma_E, \alpha^{\sigma'}_{E'}\}=0,
\ee
\be \label{rel_antibeta}
\{ \beta ^\sigma_E, \beta^{\dagger\sigma'}_{E'}\}=\delta_{\sigma,\sigma'}\delta(E-E'), \quad \{ \beta ^\sigma_E, \beta^{\sigma'}_{E'}\}=0.
\ee
The post-quench Hamiltonian is
\be
H =\sum_ {s=1,2} \left(\int_{\mathcal{E}^s_>} dE \, E \,a^{(s) \dagger}_E a^{(s)}_E - \int_{\mathcal{E}^s_<} dE \, E \,b^{(s) \dagger}_E b^{(s)}_E \right),
\ee
where
\be \label{rel_antialpha_postq}
\{ a ^s_E, a^{\dagger s'}_{E'}\}=\delta_{s,s'}\delta(E-E'), \quad \{ a^s_E, a^{s'}_{E'}\}=0,
\ee
\be \label{rel_antibeta_postq}
\{ b ^s_E, b^{\dagger s'}_{E'}\}=\delta_{s,s'}\delta(E-E'), \quad \{ b ^s _E, b^{s'}_{E'}\}=0.
\ee

An excitation that corresponds to a particle occupying an eigenstate $v_s(E;x)$ with energy $E>E_F=0$ is associated with a creation operator $a^{(s)\dagger}_E$, while an excitation that corresponds to an unoccupied eigenstate $v_s(E;x)$ with energy $E\leq E_F=0$ is a hole excitation and is associated with a hole creation operator $b^{(s) \dagger}_E$. In the above, $\mathcal{E}_>^s$ and $\mathcal{E}_<^s$ denote the energy ranges over which there exist solutions of type $s=1,2$, above or below the considered ground state level $E_F$ respectively. More explicitly, as can be seen in Fig.~\ref{fig:fig4}, eigenstates $v_1$ exist for energies $E\in(-\infty,-m)\cup(+m,+\infty)$, while $v_2$ exist for $E\in(-\infty,V-m)\cup(V+m,+\infty)$. For energies $E\in(V-m,V+m)$ the state $v_1$ becomes totally reflecting but they are still given by the same form. Similarly, for energies $E\in(-m,+m)$ the state $v_2$ becomes totally reflecting. Lastly, for energies $E\in(m,V-m)$ the states $v_{1,2}$ are given by (\ref{v'1}) and (\ref{v'2}). $E_F$ is chosen to be the one corresponding to the pre-quench Hamiltonian i.e. $E_F=V_0=0$. Overall we have $\mathcal{E}_>^{1} = [+m,+\infty)$, $\mathcal{E}_>^{2} = [0,V-m]\cup[V+m,+\infty) $, $\mathcal{E}_<^{1} = (-\infty,-m]$ and $ \mathcal{E}_<^{2} = (-\infty,0]$. The Klein zone will be denoted by $\mathcal{E}^{KZ}=[m,V-m]$. Similarly, we will denote by $\mathcal{E}_>^\sigma = (+m,+\infty)$ and $\mathcal{E}_<^\sigma=(-\infty,-m)$ the energy ranges over which there exist pre-quench solutions $u^\sigma$ ($\sigma=\pm$) above and below $E_F=0$.

 We are interested in the evolution of the total number of pre-quench particles, i.e. electrons, at large times after the quench, when considering the thermodynamic limit of our system ( i.e. the limit $L\to\infty$ with $L$ the length of the system). We would like to point out that, being the initial state of the whole system a pure one and the time evolution unitary, the system will always be in a pure state at any time, exhibiting quantum recurrences in its time evolution  which, in the present problem, are due to particles moving around the circumference $L$ of the system. On the other hand, first taking the thermodynamic limit and \emph{then} the long time limit, finite subsystems of the whole system can be described by a mixed state \cite{Bombelli-Sred}, whose observables usually exhibit a stationary limit \cite{bs-08, cdeo-08}. This prescription is the standard way of deriving the large time asymptotics in extended quantum systems. It is then clear that the only relevant observable to our analysis is the pre-quench number of particles, not the post-quench number of excitations, whose time evolution is trivial. \\
The evolution of the fermionic field $\psi(x,t)$ is formally given by expanding  on the post-quench creation and annihilation operators.

\begin{align}
& \psi(x,t) = \nonumber \\
& \sum_{s=1,2} \left[\int\limits_{\mathcal{E}_>^s}dE \; v_s(E;x) a^{(s)}_E + \int\limits_{\mathcal{E}_<^s}dE \; v_s(E;x) b^{(s)\dagger}_{E} \right] e^{-iEt} .
\end{align}

In order to calculate expectation values of observables after the quench, we need to know the initial expectation values of the post-quench creation and annihilation operators and these can be found expressing them in terms of pre-quench creation and annihilation operators, by comparison of the expansion of the field $\psi(x,0)$ in the two different bases. In the pre-quench basis we have
\begin{align}
& \psi(x,0) = \nonumber \\
& \sum_{\sigma=\pm} \left[\int\limits_{m}^{+\infty}dE \; u^\sigma(E;x) \alpha^{\sigma}_E + \int\limits_{-\infty}^{-m}dE \; u^\sigma(E;x) \beta^{\sigma\dagger}_{E}  \right] = \nonumber \\
& \sum_{\sigma=\pm} \int\limits dE \; u^\sigma(E;x) \gamma^{\sigma}_E .
\end{align}
where we defined the operator
\be
\gamma^\sigma_E \equiv
\begin{cases}
\alpha^{\sigma}_E \;  &\text{ if } E\geq m, \\
\beta^{\sigma\dagger}_E \;  &\text{ if } E\leq -m, \\
0 &\text{ otherwise. }
\end{cases} 
\ee
Clearly, $\alpha^\sigma_E$ and $\beta^{\sigma \dagger}_E$ are the electron annihilation operator and positron creation operator respectively.\\ 
Using the orthonolmalization relations, we find that
\begin{align}
\int\limits_{-\infty}^{+\infty} dx \; \left(u^\sigma(E;x)\right)^\dagger \psi(x,0) = \gamma^\sigma_E . \label{gamma}
\end{align}

Similarly, in the post-quench basis we have
\begin{align}
& \psi(x,0) = \nonumber \\
& \sum_{s=1,2} \left[\int\limits_{\mathcal{E}_>^s}dE \; v_s(E;x) a^{(s)}_E + \int\limits_{\mathcal{E}_<^s}dE \; v_s(E;x) b^{(s)\dagger}_{E}  \right] = \nonumber \\
& \sum_{s=1,2} \int\limits dE \; v_s(E;x) c^{(s)}_E . \label{c1}
\end{align}
where 
\be
c^{(s)}_E \equiv
\begin{cases}
a^{(s)}_E \;  &\text{ if } E\in\mathcal{E}_>^s, \\
b^{(s)\dagger}_E \;  &\text{ if } E\in\mathcal{E}_<^s, \\
0 &\text{ otherwise. }
\end{cases} 
\ee
The orthonolmalization relations lead now to
\begin{align}
\int\limits_{-\infty}^{+\infty} dx \; \left(v_s(E;x)\right)^\dagger \psi(x,0) = c^{(s)}_E ,
\end{align}

From (\ref{gamma}) and (\ref{c1}) we can write the pre-quench annihilation and creation operators in terms of the post-quench ones\be
\gamma^\sigma_E  = \sum_{s=1,2} \int\limits_{\mathcal{E}^s}dE' \; W_{\sigma s}(E,E') c^{(s)}_{E'} \label{transfm}
\ee
where 
\be
W_{\sigma s}(E,E') \equiv 
 \int\limits_{-\infty}^{+\infty}dx \; u^{\sigma\dagger}(E;x)  v_s(E';x) 
\ee
are the overlaps between pre-quench and post-quench eigenstates. For convenience and brevity, we will incorporate the energy zone limits into the expressions for the overlaps, defining
\be
w_{\sigma s}(E,E') \equiv W_{\sigma s}(E,E') \vartheta_\mathcal{E^\sigma}(E) \vartheta_{\mathcal{E}^s}(E')
\ee
where the function $\vartheta_\mathcal{E}(E)$ equals unit if $E\in \mathcal{E}$ and zero otherwise. 

Similarly the inverse of (\ref{transfm}) is
\be
c^{(s)}_{E'}  = \sum_{\sigma=\pm} \int\limits_{\mathcal{E}^\sigma}dE \; W^*_{\sigma s}(E,E') \gamma^{\sigma}_{E} ,
\label{transfm2}
\ee
and expresses the post-quench operators in terms of the pre-quench ones.

\subsection{Overlaps between pre-quench and post-quench eigenstates}\label{sec:overlaps}

 We will now explore the properties of the overlaps $W_{\sigma s}(E,E')$ that will be essential for the subsequent calculation. Substituting the expressions for the eigenstates and taking into account that 
\be
\int\limits_{-\infty}^{+\infty} dx \; e^{ikx} \Theta(x) = \lim_{\epsilon\to 0^+} \frac{i}{k+i\epsilon}, \label{Theta}
\ee
we find that the overlap functions have poles whenever the two energies match so as $k_{E}=k_{E'}\pm i\epsilon$ or $k_{E}=-k_{E'} \pm i \epsilon$. Explicitly, $W_{\pm 1} (E,E')$ for $E'$ outside the Klein zone is given by 
\begin{align}
W_{\pm 1} &(E,E') = \frac{i D^{\pm}(E,E')}{\pm k_{E}-k_{E'}+i\epsilon} +
\nonumber \\ & 
 \frac{i D^{\mp}(E,E') A_1(E')}{\pm k_{E}+k_{E'}+i\epsilon}
+ \frac{i D^{\pm}(E,E'-V) B_1(E')}{\mp k_{E}+k_{E'-V}+i\epsilon},
\end{align}
and for $E'$ in the Klein zone
\begin{align}
W_{\pm 1} &(E,E') = \frac{i D^{\pm}(E,E')}{\pm k_{E}-k_{E'}+i\epsilon} +
\nonumber \\ & 
 \frac{i D^{\mp}(E,E') A_1(E')}{\pm k_{E}+k_{E'}+i\epsilon}
+ \frac{i D^{\mp}(E,E'-V) B_1(E')}{\mp k_{E}-k_{E'-V}+i\epsilon},
\end{align}
where 
\begin{align}
& D^{\pm}(E,E') = \nonumber \\
& \frac1{4\pi}  \left[ \sqrt{\frac{E+m}{k_{E}}}^* \sqrt{\frac{E'+m}{k_{E'}}} \pm \sqrt{\frac{k_{E}}{E+m}}^* \sqrt{\frac{k_{E'}}{E'+m}} \right] .
\end{align}

Similarly the overlaps $W_{\pm 2} (E,E')$ for $E'$ outside the Klein zone are
\begin{align}
W_{\pm 2} &(E,E') = \frac{i D^{\mp}(E,E'-V)}{\mp k_{E}-k_{E'-V}+i\epsilon} -
\nonumber \\ & 
+ \frac{i D^{\pm}(E,E'-V) A_2(E')}{\mp k_{E}+k_{E'-V}+i\epsilon}
+ \frac{i D^{\mp}(E,E') B_2(E')}{\pm k_{E}+k_{E'}+i\epsilon},
\end{align}
and for $E'$ in the Klein zone 
\begin{align}
W_{\pm 2} &(E,E') = \frac{i D^{\pm}(E,E'-V)}{\mp k_{E}+k_{E'-V}+i\epsilon} -
\nonumber \\ & 
+ \frac{i D^{\mp}(E,E'-V) A_2(E')}{\mp k_{E}-k_{E'-V}+i\epsilon}
+ \frac{i D^{\mp}(E,E') B_2(E')}{\pm k_{E}+k_{E'}+i\epsilon}.
\end{align}

Written as functions of the energy $E$, the overlaps have simple poles close to the real axis at $E\approx E' $ and $E\approx E'-V$. All other poles and branch-cut singularities that are located away from the real $E$-axis do not matter in the thermodynamic limit, since their contribution is exponentially suppressed in the $L\rightarrow \infty$, as can be seen from the fact that $\epsilon\simeq 1/L$ (see Appendix~\ref{app:2}). In order to express the overlaps around the above poles,  it is sufficient to substitute 
\begin{align}
& \frac{i}{ k_{E}-k_{E'}+i\epsilon} = 
\frac{i \rho^{-1}(E')}{\sigma_{E'}(E-E')+i\epsilon \rho^{-1}(E')} + \nonumber \\
& \quad \frac{i \rho^{-1}(E')}{-\sigma_{E'}(E+E')+i\epsilon \rho^{-1}(E') } + ... , \; \text{ for real } k_E,k_{E'},
\end{align}
where $\sigma_E$ is the sign of $E$, $\rho(E)=|E/k_E|$ is the density of states at the energy $E$ and the dots ``...'' denote corrections that are functions non-singular along the real $E$-axis. Terms proportional to ${i}/{(k_{E}+k_{E'}+i\epsilon)}$, which do not have poles close to this axis can be omitted. Moreover, it turns out that there are no poles at opposite energies $E\approx -E'$ or $E\approx -(E'-V)$, because the corresponding residues are proportional to $D^{+}(E,-E)=0$. This expresses the fact that particles and antiparticles with the same absolute value of energy are orthogonal. The residues of the poles at $E\approx E'$ or $E\approx E'-V$ are proportional to $D^{+}(E,E)=\rho(E)/(2\pi)$ times the amplitude of the incoming, reflected or transmitted wave. Terms proportional to ${i}/{(k_{E}\pm k_{E'}+i\epsilon)}$ with $E'\in(-m,+m)$ which correspond to energies in the totally reflecting zones, can be omitted since $k_{E'}$ is imaginary in this energy window, so it cannot match with $\mp k_E$ which is always real. 

After some algebra, and taking the above substitution rules into account, we find that in all but the Klein zone, the overlaps can be written in the form 
\begin{align}
 W_{+,1} &(E,E') = \frac{i/({2\pi})}{\sigma_{E}(E-E')+i\epsilon/\rho(E)} + \nonumber \\ & \qquad \qquad 
 \frac{i B_1(E+V)/({2\pi})}{-\sigma_{E}(E-E'+V)+i\epsilon/\rho(E)} +..., \\
 W_{-,1} &(E,E') = \frac{i A_1(E)/({2\pi})}{-\sigma_{E}(E-E')+i\epsilon/\rho(E)} +..., \\
 W_{+,2} &(E,E') = \frac{i A_2(E+V)/({2\pi})}{-\sigma_{E}(E-E'+V)+i\epsilon/\rho(E)} +..., 
\end{align}
\begin{align}
 W_{-,2} &(E,E') = \frac{i/({2\pi})}{\sigma_{E}(E-E'+V)+i\epsilon/\rho(E)} +  \nonumber \\ & \qquad \qquad 
 \frac{i B_2(E)/({2\pi})}{-\sigma_{E}(E-E')+i\epsilon/\rho(E)} +..., 
\end{align}
while in the Klein zone
\begin{align}
 W_{+,1} &(E,E') = \frac{i/({2\pi})}{\sigma_{E}(E-E')+i\epsilon/\rho(E)} +...,  \\
 W_{-,1} &(E,E') = \frac{i A'_1(E)/({2\pi})}{-\sigma_{E}(E-E')+i\epsilon/\rho(E)} +  \nonumber \\ & \qquad \qquad 
 \frac{i B'_1(E+V)/({2\pi})}{\sigma_{E}(E-E'+V)+i\epsilon/\rho(E)}+..., \\
 W_{+,2} &(E,E') = \frac{i/({2\pi})}{-\sigma_{E}(E-E'+V)+i\epsilon/\rho(E)} +...,  \\
 W_{-,2} &(E,E') = \frac{i A'_2(E+V)/({2\pi})}{\sigma_{E}(E-E'+V)+i\epsilon/\rho(E)} +  \nonumber \\ & \qquad \qquad 
 \frac{i B'_2(E)/({2\pi})}{-\sigma_{E}(E-E')+i\epsilon/\rho(E)} +..., 
 \end{align}
where, as above, ``...'' denote corrections that do not involve any poles on or close to the real $E$-axis and primed quantities have been used to emphasise that we refer to the Klein zone expressions. Note that the residues $A_i,B_i$ and $A'_i,B'_i$ are smooth and bounded functions (except at the edges of the energy zones, where they are still bounded, but discontinuous or non-smooth). 
The above poles express the resonance that occurs when the incoming, reflected or transmitted wave of the post-quench eigenstate has the same energy and direction as the pre-quench eigenstate wave and is of the same type (i.e. particle or antiparticle).

\subsection{Evolution of the number of particles}

We now proceed to the calculation of the evolution of physical observables. We will focus on the total number of pre-quench particles $N(t) = \sum_{\sigma=\pm} \int_m^\infty dE \; \langle \Phi(t) | \gamma^{\sigma \dagger}_E \gamma^\sigma_E  | \Phi(t) \rangle$ 
and, in particular, its behaviour in the large time and thermodynamic limit. 
\begin{widetext}
\begin{align}
& N(t) = \sum_{\sigma=\pm} \int\limits_m^\infty dE \; \langle \Phi(t) | \gamma^{\sigma \dagger}_E \gamma^\sigma_E | \Phi(t) \rangle = \nonumber \\
& \sum_{\sigma,s_1,s_2} \int\limits_m^\infty dE \int dE_1 \, dE_2  \;  w_{\sigma s_1}^*(E,E_1) w_{\sigma s_2}(E,E_2)
\; \langle \Phi(0) | c^{s_1 \dagger}_{E_1} c^{s_2}_{E_2} | \Phi(0) \rangle \;  e^{i(E_1-E_2)t} = \nonumber \\
& \sum_{\sigma,s_1,s_2,\sigma'} \int\limits_m^\infty  dE \int dE_1 \, dE_2 \, dE' \;  w_{\sigma s_1}^*(E,E_1) w_{\sigma s_2}(E,E_2)
 w_{\sigma' s_1}(E',E_1) w_{\sigma' s_2}^*(E',E_2)  
\;  \langle \Phi(0) | \gamma^{\sigma' \dagger}_{E'} \gamma^{\sigma'}_{E'} | \Phi(0) \rangle \;  e^{i(E_1-E_2)t}
 \label{eq:N}
\end{align}
\end{widetext}
where 
\begin{align}
 \langle \Phi(0) | & \gamma^{\sigma\dagger}_{E} \gamma^{\sigma}_{E} | \Phi(0) \rangle = \nonumber \\
& \Theta(E-m)\langle \Phi(0) | \alpha^{\sigma \dagger}_{E} \alpha^{\sigma}_{E} | \Phi(0) \rangle  +  \nonumber \\
& \Theta(-E-m) \langle \Phi(0) | \beta^{\sigma}_{E} \beta^{\sigma \dagger}_{E} | \Phi(0) \rangle
\end{align}
is the particle occupation number in the initial state. 


In the present problem, in particular, the initial state is the ground state of the pre-quench hamiltonian, i.e. $\alpha^\sigma_E| \Phi(0) \rangle=\beta^\sigma_E| \Phi(0) \rangle=0$ and therefore 
\be
\langle \Phi(0) | \alpha^{\sigma \dagger}_{E} \alpha^{\sigma}_{E} | \Phi(0) \rangle = 0,
\ee 
and 
\be 
\langle \Phi(0) | \beta^{\sigma}_{E} \beta^{\sigma \dagger}_{E} | \Phi(0) \rangle = 1.
\ee
Substituting this initial density of occupied energy levels, we find
\begin{widetext}
\begin{align}
& N(t) = \sum_{\sigma,s_1,s_2,\sigma'} \int\limits_m^\infty  dE \int dE_1 \, dE_2 \int\limits_{-\infty}^{-m} dE' \;  w_{\sigma s_1}^*(E,E_1) w_{\sigma s_2}(E,E_2) w_{\sigma' s_1}(E',E_1) w_{\sigma' s_2}^*(E',E_2)  \;  e^{i(E_1-E_2)t} .
 \label{eq:N2}
\end{align}
\end{widetext}

It is convenient to consider also the long time average of the total number of particles $ \lim_{t\to\infty} \bar N (t) \equiv \lim_{t\to\infty} t^{-1} \int_0^t dt' \; N(t')$. Time averaging eliminates possible persistent oscillations and makes analytical treatment technically easier, while keeping the qualitative behaviour at large times the same, apart from such oscillations. $\bar N(t)$ is given by the same expressions (\ref{eq:N}) and (\ref{eq:N2}) but with $e^{i(E_1-E_2)t}$ replaced by 
\be
t^{-1} \int_0^t dt' e^{i(E_1-E_2)t'} = \frac{e^{i(E_1-E_2)t}-1}{i(E_1-E_2)t}.
\ee 

\begin{figure}[h!]
\begin{center}
\includegraphics[clip,width=0.9\linewidth]{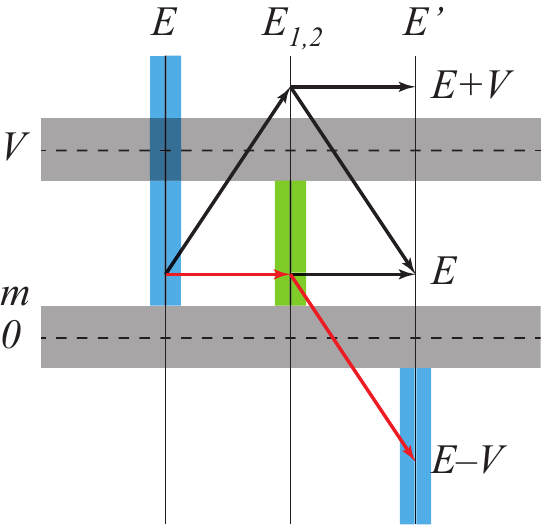}
\caption{\label{fig:resonance} Resonance between energies in the integral (\ref{eq:N}). $E$ and $E'$ match with $E_{1,2}$ in four possible ways, indicated with arrows. Since they are integrated over the windows shaded in blue, the only allowed possibility is $E\approx E_1\approx E_2 \approx E'+V$, indicated with red arrows. Then $E_{1,2}$ are restricted to values within the Klein zone, shaded green. The grey energy zones $(-m,+m)$ and $(V-m,V+m)$ indicate the totally reflecting zones.}
\end{center}
\end{figure}

In the thermodynamic and large time limit, the above expressions are dominated by the poles of the overlaps that are close (at distance $\epsilon\sim 1/L$) to the real $E$ and $E'$ axes and within the energy windows under integration. Since each of the overlap factors in the integrand $w_{\sigma s}(E,E')$ is highly peaked around the two resonance poles, at $E\approx E'$ and $E\approx E'-V$, and the time averaging factor is highly peaked at $E_1\approx E_2$, there are four possibilities to match the four energies $E,E_1,E_2$ and $E'$: $E_1$ matches with $E_2$, while $E$ and $E'$ independently match with either $E_{1,2}$ or $E_{1,2}-V$ (i.e. $E'$ matches with either $E+V$ or $E$ or $E-V$). But integration over $E$ and $E'$ is restricted to the windows $(+m,+\infty)$ and $(-\infty,-m)$ respectively, so there remains only one possibility that leads to overlapping energy windows: $E\approx E_1\approx E_2 \approx E'+V$. This possibility exists only for $V>2m$ and constraints the variables $E_1\approx E_2\approx E$ to be within the Klein zone while $E' \in (m-V,-m)$  (Fig.~\ref{fig:resonance}). Therefore, using this approach, the expression for $\bar N(t)$ reduces to a single energy integral over the Klein zone only (see Appendix~\ref{app:1} for details). This clearly shows that,  in the thermodynamic and large time limit there is no steady-state particle production for $V<2m$, i.e. for the values of potential which do not admit the existence of the Klein Zone.

Substituting the expressions for the resonance poles of the overlaps, using the properties of the coefficients $A_{i}$ and $B_{i}$ and performing the integration using the residue theorem (Appendix~\ref{app:1}), we finally find the simple expression 
\be
\bar N(t)= \frac{1}{4\pi} t \int \limits_{m}^{V-m} T(E) dE 
\label{result0}
\ee
where the transmission coefficient $T(E)$ in the Klein zone is given by (\ref{T'}). 

\begin{figure}[ht]
\begin{center}
\includegraphics[clip,width=1.0\linewidth]{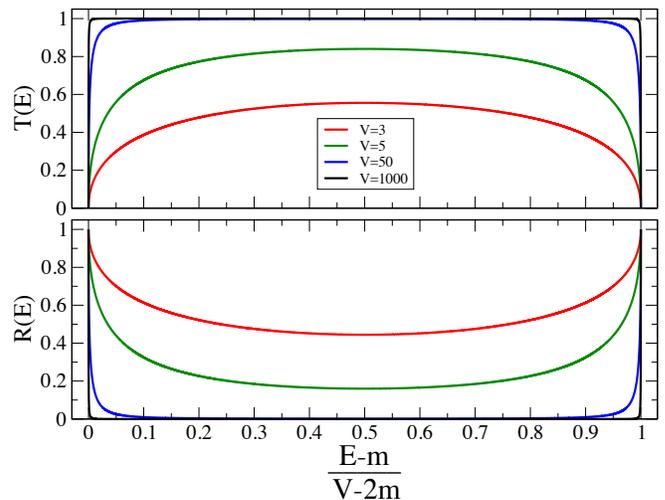}
\caption{\label{fig:plot_Te_Re_1} The transmission coefficient $T(E)$  and the reflection coefficient $R(E)$ in the Klein zone $(m,V-m)$, as a function of the rescaled energy $(E-m)/(V-2m)$, for several values of $V/m$. Note that $T(E)\to 1$, $R(E)\to 0$ for $E=V/2$, as $V\to\infty$. }
\end{center}
\end{figure}

\begin{figure}[ht]
\begin{center}
\includegraphics[clip,width=1.0\linewidth]{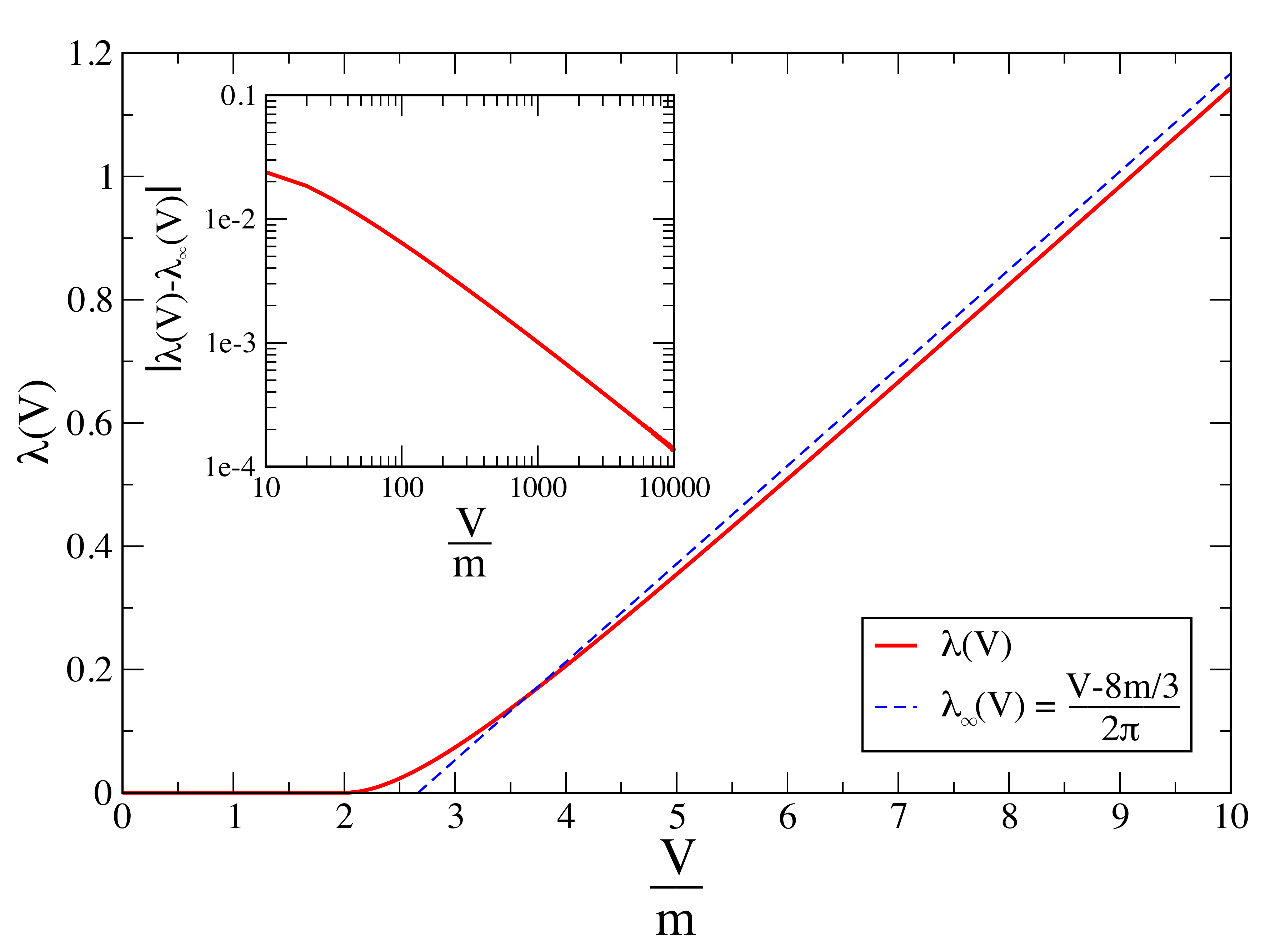}
\caption{\label{fig:plot_plambda} The large time rate of particle production $\lambda$ as a function of the potential $V/m$. Note that $\lambda \sim V$ as $V\to\infty$. The dashed line shows the asymptotic linear increase while in the insect the difference between the two is reported as a function of the potential.}
\end{center}
\end{figure}

Our result shows that, if $V>2m$, $\bar N(t)$  increases linearly with time at a constant rate, so the large time limit is described by a Non Equilibrium Steady State \cite{transport}.
This linear increase for $V>2m$ is obviously a direct consequence of the existence in the post-quench hamiltonian of classically forbidden scattering in the Klein energy zone. Moreover, the rate of production of particles 
\be
\lambda \equiv \lim_{t\to\infty} \frac{N(t)}{t} = 2 \lim_{t\to\infty} \frac{\bar N(t)}{t} = \frac{1}{2\pi} \int \limits_{m}^{V-m} T(E) dE , \label{result}
\ee
turns out to increase when $V$ increases, in accord with the behaviour of the transmission coefficient in the Klein zone 
(see fig.~\ref{fig:plot_Te_Re_1}). For $V$ much larger than $2m$, the particle production rate, indicated with $\lambda_{\infty}(V)$, is given by 
\be\label{rate_asymp}
\lambda_\infty(V)=\frac{V-8m /3}{2\pi}\quad\quad V\gg 2m,
\ee
which is linear in $V$. In Fig.~\ref{fig:plot_plambda} we show the particle production (continuous red line) and its asymptotic value (blue dashed one) as a function of the potential. The former is exactly vanishing for $V/m<2$ and it approaches its asymptotic value, as shown in the inset, with a precision of order $10^{-2}$ when ${V}/{m}\simeq 10$. 

Our analytic result (\ref{result}) is in agreement with recent numerical computations of the electron production rate \cite{Krekora, Cheng} in a three-dimensional system of Dirac fermions with a rounded step-like potential. The authors compute numerically the total electron population and find that, for a supercritical potential ($V>2m$), there is a linear growth at constant rate for $t\rightarrow \infty$ while for a subcritical potential ($V<2m$) there is no such growth. Even though their setup is not exactly the same as ours, the authors of \cite{Krekora} find that their numerics are described by precisely the same formula (\ref{result}) where $T(E)$ is the corresponding transmission coefficient in their setup. Moreover, it can be easily checked by a trivial generalization of our calculation that our result (\ref{result}) is still valid when the initial value $V_0$ of the homogeneous potential is not zero but between zero and $V$.

\subsection{Physical interpretation}
The physical interpretation of the above results can be based on a semiclassical approach \cite{CC,semicl_ex}, which has proved to give correct results in many cases of quenches including in inhomogeneous problems. In the semiclassical approximation we view quantum excitations as quasiparticles with specified position and momentum that move ballistically with a velocity given by the group velocity that corresponds to their momentum $v_g\leq c=1$. A global quench injects an extensive amount of energy into the system by creating such quasiparticle excitations. 
These quasiparticle excitations are created at the time of quench, typically in pairs of coherent quasiparticles with opposite momenta (in the case of non-interacting systems). In the thermodynamic and large time limit where dephasing has typically eliminated quantum coherence effects, physical observables can be expressed as classical probability averages of the contributions of all such quasiparticles. This approximation is also valid for any time and system size, in the limit of low densities where coherence effects are unimportant.

It is instructive to analyse, at the semiclassical level, what happens in a homogeneous quench of the potential from zero value to $V$.  In this case the Fermi level is shifted by a constant value equal to the height $V$ of the potential, so that in comparison with the post-quench ground state, the (filled) pre-quench Fermi sea would correspond to a depleted or raised Fermi sea, depending on the sign of $V$. From the structure of the overlaps in the homogeneous case we deduce that a pre-quench creation or annihilation operator can be expressed in terms of post-quench operators with the same momentum; hence the semiclassical picture that emerges here is that the initial state is a source of non-paired quasi-particles. We highlight that this is slightly different from the typical semiclassical picture where the initial state can be seen as source of oppositely moving pairs of coherent quasiparticles. We would like to emphasize that, in the homogeneous quench case, there is no net particle production. What happens is that, due to the shift of the potential and the consequent shift of the Fermi energy, the pre-quench excitations may have become particle or antiparticle according to the post-quench Hamiltonian.

\begin{figure}[ht]
\begin{center}
\includegraphics[clip,width=0.9\linewidth]{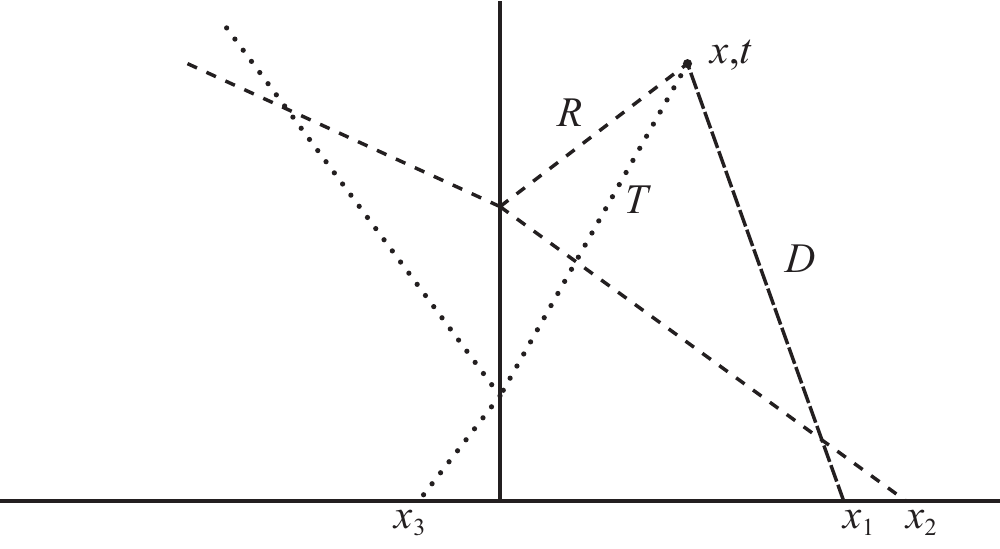}
\caption{\label{fig:Fig_Klein_Paradox_1}Space-time diagram representing the three possible origin for a quasi-particle passing at a point $(x,t)$ in the semiclassical picture.  }
\end{center}
\end{figure}

This picture can be applied also in the case of a step-like profile of the potential, even though in this case the initial quasiparticle density is inhomogeneous. The situation is depicted in  Fig.~\ref{fig:Fig_Klein_Paradox_1}, where it is shown that the particle density $\rho(x, t)$ is given by the sum of three contributions: direct particles (D) that have not passed from the origin before arriving at the space-time point $(x, t)$, reflected particles (R) that have been reflected at the origin and transmitted particles (T) that have been transmitted/refracted through the origin. 
To obtain the long time behaviour of observables, we can neglect the contribution of the quasiparticles produced in the quench from the region near the origin and consider just those produced from the regions at $x \to \pm \infty$.  In these regions, where the potential is spatially constant, the initial quasiparticle production is that corresponding to an homogeneous quench to the local value of the potential. After the quench the quasiparticles travel ballistically with group velocity $v_g$ with $v_g\leq c$ and, when they arrive at the origin, they scatter with the potential step. It is the scattering of the quasiparticles produced after the quench with the potential step that allows  net particle production. If $V>2m$ the scattering of antiparticles incoming from the right produces particles with probability $T(E)$. 
Therefore the number of particles increases with time and more precisely it increases linearly, since for large times the number of incoming excitations reaching the origin is the number of all initial excitations created at large distances. 
We conclude that the semiclassical approach correctly describes the physics of the constant rate of  particle production connected to  Klein tunneling.

\section{Conclusions}
In this work we study the Klein tunneling by performing an inhomogeneous quantum quench of the potential in a system of relativistic one dimensional fermions. We determine the full energy spectrum of the inhomogeneous system and explain the necessity for a special definition of the eigenstates in the Klein zone due to the orthonormalization requirement. We further derive the overlaps of these eigenstates with the homogeneous ones, which allows us to study the evolution of physical observables in the thermodynamic and large time limit, exploiting the resonances of pre-quench with post-quench eigenstates. In particular, we derive analytically the particle production rate $\lambda$ and show that its dependence on the ratio $V/m$ reflects the paradoxical fact that the transmission of relativistic fermions inside a potential step is nearly complete, when the step size is large.  We  have therefore demonstrated that a quantum quench of the potential can be used to reveal the presence of Klein tunneling, especially since the extensive amount of energy injected by the quench leads to a steady particle production.    

Our method shows that the resonance transitions are dominant in the thermodynamic and large time limit. From this point of view, it is reminiscent of the semiclassical theory of radiation and Fermi's Golden Rule, even though the latter relies also on perturbation theory, while our method does not. This explains why our results are consistent with the semiclassical interpretation of quantum quenches, as we also showed in more detail above. Following this observation, we expect our result in the form (\ref{result}) to be valid even in different spatial dimensions and for different profiles of the potential step. In fact, as we mentioned, (\ref{result}) is in perfect agreement with the formula that was found to describe numerics in a three-dimensional analogue of our problem with a smoothed step \cite{Krekora}.

Although the  major challenge of the steepness of the potential is still present even in our quench setting, we have analytically shown that the physics of the Klein tunnelling can be obtained in a context very different from the usual one. The novelty of our approach consists in the fact that by quenching the value of the potential we no longer need to have an incident beam of particles. A possible experimental realization could involve  highly controllable systems in table-top experiment with optical lattices and ultracold atoms.

It would be interesting to investigate the consequences of quenching other parameters of the Dirac equation, as for example the mass or the mass phase of the relativistic Dirac fermions from a constant value to an inhomogeneous kink like shape \cite{Zar}. Actually vacuum polarization effects involving fermionic fields interacting with background solitons have been shown to induce fractional fermion number localized on the soliton \cite{fract_charge}. Following the stimulating proposals of \cite{fract_real_olattice} and \cite{angelakis_2014} it would be interesting  to observe the consequences of a quantum quench on the charge fractionalization mechanism.

\section*{Acknowledgments}

The work of S.S. was supported by the ERC under Starting Grant 279391 EDEQS. We thank Pasquale Calabrese, Andrea Trombettoni and Andrea De Luca for useful discussions and insights.

\onecolumngrid


\appendix

\section{Finite volume energy eigenstates}\label{app:2}

As typically in quantum field theory, infrared divergences occurring in observables can be resolved by setting up the problem on a finite size system. Momentum or energy integrals are then replaced by sums over discrete sets of eigenvalues, determined by the quantization conditions, i.e. the boundary conditions of the finite size system, which are assumed periodic. In the present problem this means that in determining the energy eigenstates, apart from the matching condition at the origin, we also have one at $x=\pm L/2$, where $L$ is the length of the system. Following the same procedure as before, we write the eigenstates in the general form
\begin{align}
& \psi(x,t) =
\begin{cases} 
A u^+_0(E;x) + B u^-_0(E;x), & \quad x<0\\
C u^+_V(E;x) + D u^-_V(E;x) , & \quad x>0 \\
\end{cases},
\end{align}
and impose the conditions 
\begin{align}
\psi(x\to0^+,t) & =\psi(x\to0^-,t), \\
\psi(x\to -L/2,t) & =\psi(x\to +L/2,t),
\end{align}
from which, eliminating the coefficients $A,B,C$ and $D$, we find that the wavenumbers $p=k_{E}$ and $q=k_{E-V}$ must also satisfy the condition
\be
\frac{\sin^2[(p+q)L/4]}{\sin^2[(p-q)L/4]} = \left(\frac{1-\kappa}{1+\kappa}\right)^2,   \label{Espectrum}
\ee
with $\kappa$ given in terms of $p,q$ and $E$ in (\ref{kappa}). Expressed in terms of the only independent variable $E$, this is the equation for the energy eigenvalues at finite size $L$. Note that it is a transcendental equation and can be solved numerically, although several qualitative results about the solutions can be derived without explicitly solving it.
We notice that due to the periodicity of trigonometric functions, the distance between two successive quantized momenta is of the order $\pi/L$ 
 and therefore the density of states in momentum space is uniform. 
Other results, like the absence of solutions in the energy window $(V-m,m)$ for $V<2m$ and the degree of degeneracy of solutions, can also be concluded from the form of the equation. 

In evaluating expressions for physical observables, we can use a standard trick \cite{FVR} in order to pass from discrete energy sums to integrals in the thermodynamic limit. The energy sum can be written as a sum of residues of a complex function with single poles at the roots of the equation (\ref{Espectrum}). Next the sum of residues can be written as a contour integral, more specifically a sum of two integrals, one above and one below the real energy axis, excluding any other possible poles that are on or close to the real axis (at distance $\sim 1/L$). In the thermodynamic limit, one of these integrals vanishes because the integration contour can be deformed to imaginary infinity where it decays typically exponentially with $L$. If there are any singularities away from the real axis that are crossed while deforming the contour, their contribution vanishes also exponentially with $L$. The other of the two integrals gives the desired thermodynamic expression for the observable, along with the correct prescription of how the contour passes around any pole that is on or close to the real axis. 

Applying this method in our present problem, firstly justifies the regularization of infrared divergences with $\epsilon\sim 1/L$ used in the main text, since this is the order of the distance between successive energy eigenvalues. Secondly it justifies that any poles whose distance from the real axis is of order more than $\epsilon$ do not matter in the thermodynamic limit and in fact they typically decay exponentially with $L$.

\section{Evaluation of energy integrals}\label{app:1}

We will show how the asymptotics of the multiple energy integrals in (\ref{eq:N}) in the thermodynamic and large time limit, can be derived by deformation of the integration contours in the complex energy plane and using the analyticity properties of the integrated function, more specifically the resonance poles of the overlaps which we studied in Section \ref{sec:overlaps}. The following analysis would give the exact value of the multiple energy integral for all times and for large but finite system sizes, if these poles were the only singularities of the overlaps in the complex energy plane. This is not true, first because the overlaps are piecewise functions, i.e. they have a different expression in the Klein zone in comparison with the others, and second because the functions $D^{\pm}(E,E')$, $A(E)$ and $B(E)$ exhibit branch cuts and poles away from the real axis. These non-analyticities prevent the deformation of the integration contours in the complex energy plane. However, given that we are only interested in the limit $\epsilon\sim 1/L\to 0$ and $t\to\infty$, the asymptotic form of the integrals is given by the contribution of the resonance poles and therefore it is sufficient to know that $A(E)$ and $B(E)$ are simply bounded functions for all energy values and smooth at the poles. Indeed, as argued in Appendix~\ref{app:2}, the contribution of any singularity away from the real energy axis is suppressed in the thermodynamic limit. For the same reason the integration ranges can be extended from any energy zone to the whole real axis $(-\infty,+\infty)$, provided this does not introduce additional resonance poles, because the asymptotics are not affected by this extension. It should also be noted that the piecewise form of the overlaps can be understood by expressing the eigenstates as functions that are analytical in a non-trivial Riemann surface with suitably chosen branch-cuts, based on physical requirements \cite{Bosanac}. This suggests that it may actually be possible to deform the integration contours, so that our resonance pole approach that we present below can be extended to give exact results for any time, not only at large times. Such an exhaustive analysis is however beyond the scope of our present work and it is sufficient for us to consider this approach as a widely used approximation.

The first step in evaluating the integral is to substitute the expressions for the overlaps keeping only the relevant energy poles, i.e. those close to the real energy axes. These poles lead to matching of the energies in pairs, so the next step is to identify all possible energy pairing possibilities that are consistent with the ranges of the energy zones under integration. We thus find $E\approx E_1\approx E_2 \approx E'+V$, so that we can replace all energy variables by a single variable everywhere except at the poles, since the positions of the poles and the prescription for the integration around them, determines the asymptotic $L$ and $t$ behaviour. These substitutions are essentially the outcome of evaluating the residues of the poles and can be performed directly. We finally find three types of integrals: 
\begin{align}
I_1= & \int \frac{dE dE_1 dE_2 dE'}{(2\pi)^4} \; F_1(E) \times \nonumber \\
&\frac{e^{i(E_1-E_2)t}-1}{i(E_1-E_2)t}
\frac{i}{(E-E_1+i\epsilon\rho_E^{-1})}  \frac{i}{(E_2-E+i\epsilon\rho_E^{-1})}  \frac{i}{(E'+V-E_1+i\epsilon\rho_{E-V}^{-1})}  \frac{i}{(E_2-E'-V+i\epsilon\rho_{E-V}^{-1})} \\
I_2= & \int \frac{dE dE_1 dE_2 dE'}{(2\pi)^4} \; F_2(E) \times \nonumber \\
&\frac{e^{i(E_1-E_2)t}-1}{i(E_1-E_2)t}
\frac{i}{(E-E_1+i\epsilon\rho_E^{-1})}  \frac{i}{(E_2-E+i\epsilon\rho_E^{-1})}  \frac{i}{(E_1-E'-V+i\epsilon\rho_{E-V}^{-1})}  \frac{i}{(E'+V-E_2+i\epsilon\rho_{E-V}^{-1})} \\
I_3= & \int \frac{dE dE_1 dE_2 dE'}{(2\pi)^4} \; F_3(E) \times \nonumber \\
&\frac{e^{i(E_1-E_2)t}-1}{i(E_1-E_2)t}
\frac{i}{(E_1-E+i\epsilon\rho_E^{-1})}  \frac{i}{(E-E_2+i\epsilon\rho_E^{-1})}  \frac{i}{(E_1-E'-V+i\epsilon\rho_{E-V}^{-1})}  \frac{i}{(E'+V-E_2+i\epsilon\rho_{E-V}^{-1})}
\end{align}

Each of the three integrals can be evaluated using the residue theorem. First we perform the integration over $E$ and then over $E'$, closing the integration contour either above or below the real axis.Then we can perform the integration over $E_1$ using the residue theorem in the form
\be
P.V. \int \limits_{-\infty}^{+\infty} dz \; f(z) = 2 i \pi \left( \sum_{\text{poles} \atop {\text{in UHP}}} {\text{Res}} f(z) + \frac12 \sum_{{\text{poles on} \atop \text{real axis}}} 
{\text{Res}} f(z)  \right ),
\ee
which is valid for any function $f(z)$ that decays sufficiently fast for $z\to i \infty$, as in the present case, where the integrand decays exponentially as $E_1\to i\infty$. The remaining integration cannot be performed without explicit knowledge of the functions $F_i$. However, provided that convergence allows it, we can postpone the integration for later and take first the thermodynamic limit $\epsilon\to 0$ followed by the large time limit $T\to\infty$ to find the asymptotic behaviour we are looking for. 

Following this procedure for each of the three types of integrals, we find for the above asymptotic limits 
\begin{align}
& I_1 = \frac12 t \int\limits_{m}^{V-m} \frac{dE}{2\pi} F_1(E) , \\
& I_2 = \int\limits_{m}^{V-m} \frac{dE}{2\pi} F_2(E) \left(\frac1\epsilon - t \rho_E^{-1} \right) \frac{1}{\rho^{-1}_E + \rho^{-1}_{E-V}} , \\
& I_3 = 0.
\end{align}

The explicit expressions of the functions $F_1(E)$ and $F_2(E)$ are 
\begin{align}
& F_1(E) = |B_2(E)|^2 = T(E) , \\
& F_2(E) = \left|A_1(E) B_1^*(E) + A_2^*(E) B_2(E)\right|^2 = 0 ,
\end{align}
where the coefficients $A_i$ and $B_i$ are given by (\ref{AB'}), since the remaining energy integration is over the Klein zone. Summing up all terms, we find our final result (\ref{result0})
\be
\bar N(t)= \frac1{4\pi} t \int \limits_{m}^{V-m} T(E) \, dE.
\ee

\end{document}